\documentclass[11pt,a4paper,english,nofootinbib,,superscriptaddress]{revtex4}
\usepackage{lmodern}

\usepackage[T1]{fontenc}
\usepackage[latin9]{inputenc}
\usepackage{babel}
\usepackage{color}
\usepackage{cancel}
\usepackage{amsmath}
\usepackage{graphicx}
\usepackage{amssymb}
\usepackage{esint}
\usepackage[unicode=true, pdfusetitle,
bookmarks=true,bookmarksnumbered=false,bookmarksopen=false,
breaklinks=false,pdfborder={0 0 1},backref=false,colorlinks=false]
{hyperref}
\usepackage{amsmath}
\usepackage{amssymb}
\usepackage[compat=1.0.0]{tikz-feynman}
\usepackage{float}

\def\barra#1{\not \!#1}
\def\b{\begin{equation}} \def\e{\end{equation}}
\def\bd{\begin{displaystyle}} \def\ed{\end{displaystyle}}
\def\ba{\begin{array}} \def\ea{\end{array}}

\def\bee{\begin{enumerate}}
	\def\eee{\end{enumerate}}

\def\ud{\mathrm{d}}

\def\dg{^{\dag}}

\def\1{\mbox{I\hspace{-.15em}1}}

\def\b{\begin{equation}}
\def\e{\end{equation}}
\def\bee{\begin{enumerate}}
	\def\eee{\end{enumerate}}

\makeatletter
\usepackage{latexsym}\usepackage{bm}

\makeatother

\begin{document}
	\title{Interaction of massless minimally coupled scalar field with spinor field
		\\ in de Sitter universe}
	
	\author{Y. Ahmadi}
	\email{ahmadi.pave@gmail.com} \affiliation{Department of Physics, Amirkabir University of Technology, Tehran, Iran}
	
\begin{abstract}

\noindent \hspace{0.35cm}
 The interaction of massless minimally coupled scalar field and spinor field $ \left( \Psi + \Phi_m \rightarrow \Psi + \Phi_m \right) $ is investigated and the behavior of interaction and scattering matrix in Minkowski limit is studied. It is seen that the massless minimally coupled scalar field operator, the interaction Lagrangian and the scattering matrix become zero. Also the incoming and outgoing spinor fields, which in de Sitter space-time are different and are interacted with massless minimally coupled scalar field, in null curvature limit are the same and no interaction seen. In other word, the massless minimally coupled scalar field is dependent on curvature and can be considered as a part of a gravitational field.

\end{abstract}

\maketitle


\section{Introduction}
Constructing the quantum field theory (QFT)  in ambient space formalism of de Sitter (dS) space-time has been followed in last years \cite{ta97,77}. The quantization of scalar, spinor and vector free field is done almost completely \cite{gata, gagarota, rota05, ta01, berotata, ta96, brgamo} and spin $\frac{3}{2}$ and spin $2$ fields is investigated in last years \cite{taazba, gagata, tatafa, ta09, fatata, azam, petata, taro12, derotata}.

For interaction fields, the interaction of spinor field with vector field is investigated and in previous paper and scattering matrix of Compton scattering in ambient space formalism is calculated and also it is showed that in null curvature limit this scattering matrix is match with Minkowski counterpart \cite{jaahta2017}. Another electron-photon interaction is wave mixing which is known in optics. This interaction was investigated from QFT point of view and in ambient space formalism of dS universe \cite{photon decaying}.

The interaction between massless minimally coupled scalar field and spinor field investigated and the scalar-spinor interaction Lagrangian was obtained by representing a new transformation in way which is similar to group theoretical \cite{higgs}. The interaction of massless minimally coupled (\textit{mmc}) scalar field with spinor field was investigated and scattering matrix of $\left(\Psi+\Psi\rightarrow\Psi+\Psi\right)$ studied \cite{ahjata2019}. This interaction in flat space limit was discussed and it be seen that the \textit{mmc} scalar field, interaction Lagrangian and scattering matrix vanished in null curvature limit \cite{flat limit}.
Here, following on the interaction of \textit{mmc} scalar field with spinor field we investigate the interaction $\left(\Psi+\Phi_m\rightarrow\Psi+\Phi_m\right)$ 
 and calculate the scattering matrix and then study the behavior of this interaction in null curvature limit.
 
The organization of this article is as follow. The terminology and notation which used in this paper is presented in section \ref{notation}. The scaler-spinor interaction is investigated in dS ambient space formalism in section \ref{interaction} and in null curvature limit in section \ref{flat}. Finally, the conclusion has been presented in section \ref{Conclusion}.

\setcounter{equation}{0}	
\section{terminology and notation}\label{notation}

 The ambient space formalism is 5-dimensional Minkowski space and dS hyperboloid is 4-dimensional hyperboloid embedded in this ambient space.
 In dS ambient space formalism, the action of spinor free field and \textit{mmc} scalar free field is:
\cite{77,bagamota,higgs}:
\b \label{conf spinor massless action} S(\Psi,\Phi)=\int \ud\mu(x){\cal L}_{free}(\Psi , \Phi)=\int \ud\mu(x)\left[H \bar{\Psi} \gamma^4\left( -i\barra{x} \barra{\partial}^\top+2i\pm\nu\right) \Psi+\Phi_m \;\partial^\top\cdot\partial^\top\;\Phi_m \right],\e 
where the $d\mu(x)$ is dS-invariant volume element. Also $\barra x=\eta_{\alpha\beta}\gamma^{\alpha} x^{\beta}$ and the five matrices $\gamma^{\alpha}$, satisfy the conditions 
$
\gamma^{\alpha}\gamma^{\beta}+\gamma^{\beta}\gamma^{\alpha}
=2\eta^{\alpha\beta}$ and $\gamma^{\alpha\dagger}=\gamma^{0}\gamma^{\alpha}\gamma^{0}$ \cite{ta97,ta96,bagamota}.
It is should be noted that these ambient $\gamma$ matrices are different from Minkowski matrices $\gamma^{'}$,
\b \label{gamma relation}\gamma'^{\mu}=\gamma^{\mu}\gamma^4.\e
In equation \eqref{conf spinor massless action}, the $\nu$ is related to dS mass parameter and the $\partial^\top_\alpha=\partial_\alpha+H^2x_\alpha x\cdot\partial$ is transverse derivative on hyperboloid. The $\Psi$ is spinor field which satisfies the dS-Dirac equation and also the $\bar{\Psi}=\Psi\dg\gamma^0\gamma^4$ is the adjoint spinor field \cite{77}. In dS ambient space formalism, the spinor field operator is \cite{77}
\b \label{psi expansion}\Psi(x)\equiv \Psi^{(+)}+\Psi^{(-)}=\int_{S^3}d\mu(\xi)\sum_{n}\left[a( \tilde{\xi},n)(Hx\cdot\xi)^{-2-i\nu} {\cal U} (x,\xi,n)+b^{\dag}(\xi,n)(Hx\cdot\xi)^{-1+i\nu} {\cal V} (x,\xi,n)\right],\e
where and the $a(\tilde{\xi},n)$ and $b\dg(\xi,n)$ are the annihilation and creation operators respectively \cite{77}. Also the $\cal U$ and $\cal V$ are spinors which them explicit form are discussed in \cite{bagamota}.
The $\Phi_m$ is the \textit{mmc} scalar field which related to massless conformally coupled (\textit{mcc}) scalar field ($\Phi_c$) as  \cite{higgs,77,khrota,gareta}:
\b\label{magic}
\Phi_m(x)= \left[HA\cdot\partial^\top + 2 H^3 A\cdot x\right]\Phi_c(x),\e
where $A_\alpha$ is arbitrary constant which discussed in \cite{flat limit}.
the \textit{mmc} scalar field operator can be obtained as \cite{higgs}:
\b \label{mmc sc-fi operator}
\begin{array}{clcr}
	\displaystyle \Phi_m(x)\equiv\Phi^{(+)}+\Phi^{(-)}
	&=\displaystyle \boldsymbol{N}_{\Phi_m} \sum_{l=1}^5\int_{S^3}   d\mu(\xi) \left\lbrace\; a(\tilde{\xi},0;0,1)\left[-2 \left(H^2 A^{(l)}\cdot\xi\right)\left(Hx\cdot\xi\right)^{-3}\right] \right.
	\\
	\\
	&\left. \displaystyle +a^{\dag}(\xi,0;0,0)\left[- \left(H^2A^{(l)}\cdot\xi\right) \left(Hx\cdot\xi\right)^{-2} +\left(H^3 A^{(l)}\cdot x\right)\left(Hx\cdot\xi\right)^{-1}\right]
	\; \right\rbrace.
\end{array} \e

\section{Interaction}\label{interaction}
In \eqref{conf spinor massless action}, if the transverse derivative $\partial_\alpha^\top$ be replaced with this new derivative $D_\alpha\Psi=(\partial_\alpha^\top+{\cal G}B^\top_\alpha\Phi_m)\Psi $, the action be invariant under the following transformation \cite{higgs}:
\b\label{transformation}
\Psi\longrightarrow\Psi'=e^{-\frac{i}{2}(Hx\cdot B)^{-2}}\Psi
\;\;\;,\;\;\;
\Phi_m\longrightarrow\Phi'_m=\Phi_m+(Hx\cdot B)^{-3}\,.
\e
Then the interaction Lagrangian is obtained as \cite{higgs}:
\b \label{higgs spinor lagrangian} {\cal L}_{int}=-i{\cal G}\;H\bar{\Psi} \gamma^4 \barra{x} \barra{B^\top}\Phi_m\Psi, \e
where $B_\alpha$ is an arbitrary 5-vector constant that $B^\alpha B_\alpha=0$ and ${\cal G}$ is the coupling constant which determines the interaction intensity.

The evolution of quantum states determined by time evolution operator $\left|\alpha,t\right>=U(t,t_0)\left|\alpha,t_0\right>$ and the scattering matrix is defined as ${\cal S}\equiv U(-\infty,+\infty)\quad,\quad\left|\alpha,+\infty\right>={\cal S}\left|\alpha,-\infty\right>$. The time evolution operator can be defined in dS space time in static coordinate, but in curved space time due to the curvature and event horizon, the S-matrix can not be defined generally \cite{flilto,anilto,ta17}. Fortunately for purpose, \textit{i.e.} interaction of fields, the interaction occurred in atomic scale. In this scale, one can consider only the indirect gravitation effect and ignore the direct effect of gravitation. So the $U(t, t_0)$ can be expressed as:
$$U(t,t_0)={U_M}(t,t_0)+ Hf(t,t_0)+...\;$$
which the ${U_M}(t,t_0)$ in null curvature limit become the Minkowski time evolution operator and the $f(t,t_0)$ includes the gravitational direct effect.
Then the S-matrix in atomic scale approximation can be expressed from its Minkowski counterpart as \cite{jaahta2017,ahjata2019}:
\b \label{ds s matrix expansion}
\displaystyle{\cal S}\simeq \sum_{\lambda=0}^{\infty}{\cal S}^{(\lambda)},\;\;
\displaystyle{\cal S}^{(\lambda)}=\dfrac{(-i)^\lambda}{\lambda!}\int\ud^4x_1\cdots \int\ud^4x_\lambda\;\;T\left[{\cal H}_{int}(x_1)\;\cdots\;{\cal H}_{int}(x_\lambda)\right],
\e
where $T[\cdots]$ is time order product of fields. So the ${\cal S}^{(2)}$ is obtained as:
\b \label{s2 time order}{\cal S}^{(2)}=-\dfrac{1}{2}{\cal G}^2\int\int\ud\mu(x_1) \ud\mu(x_2)\;\;T\left[\left(\bar{\psi}\gamma^4H\barra{x}\barra{B}^\top\phi_m\psi\right)_{x_1}\left(\bar{\psi}\gamma^4H\barra{x'}\barra{B}^\top\phi_m\psi\right)_{x_2}\right].\e
The time order product of fields by using Wick's theorem, can be written in terms of the normal order product of fields. The time order product of \eqref{s2 time order}  is written as eight normal order products but the term that correspond to interaction $\Psi+\Phi_m\rightarrow\Psi+\Phi_m$ (as shown in figure \ref{fig}) is:
$$
\begin{array}{clcr}
\displaystyle {\cal S}^{(2)}
&=\displaystyle -\dfrac{{\cal G}^2}{\hslash^2}\int\int\ud\mu(x_1) \ud\mu(x_2)
\; N\left[
\bar{\Psi}(x_1)\gamma^4\;H\cancel{x}_1\;\cancel{B}^\top\Phi_m(x_1)\underleftrightarrow{\Psi(x_1)
	\;\; \bar{\Psi}}(x_2)\gamma^4\;H\cancel{x}_2\;\cancel{B}^\top\Phi_m(x_2)\Psi(x_2)
\right]
\\
\\
&=\displaystyle \dfrac{i{\cal G}^2}{\hslash^2}\int\int\ud\mu(x_1) \ud\mu(x_2)\left(\gamma^4\cancel{B}^\top H\cancel{x}_1\;\boldsymbol{S}(x_1,x_2) \gamma^4\;H\cancel{x}_2\cancel{B}^\top\right)_{ir} N\left[\bar{\Psi}_i(x_1)\Phi_m(x_1)\Phi_m(x_2)\Psi_r(x_2)\right]
\end{array}
$$
where, the $N[\cdots]$ is normal order product and $\boldsymbol{S}(x_1,x_2)=-i\left<\Omega|\Psi(x_1)\bar{\Psi}(x_2)|\Omega\right>$ is the spinor two-point function \cite{77,ta96,bagamota}. It is should be noted that, given that the $\Psi,\;\bar{\Psi},\;\barra{x},\;\barra{B}^\top$ and $\boldsymbol{S}$ have matrix representation,
they are written in terms of their component and the $i$ and $r$ indexes are matrix components but the $m$ index of $\Phi$ is symbol of \textit{mmc} scalar field operator. By separating the $\Psi,\;\bar{\Psi}$ and $\Phi_m$ in positive energy and negative energy parts one can expand the normal order product and extract the physical terms among them. So the ${\cal S}^{(2)}$ for interaction shown in figure \ref{fig}, is obtained as:

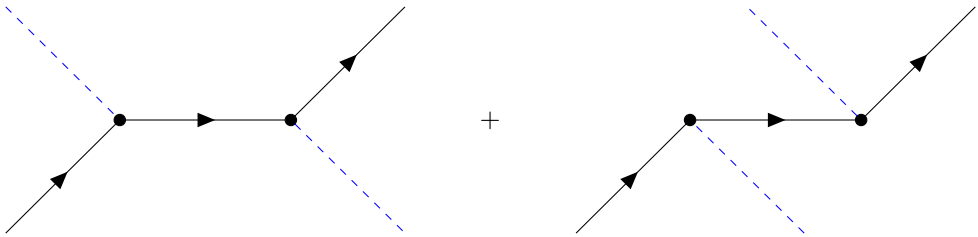
\begin{figure}[H]
	\centering
	\begin{tikzpicture}[scale=1.5]
	\begin{feynman}
	\vertex (1) at(0,-1);
	\vertex (2) at(0,1);
	\vertex [dot] (3) at(1,0){} [];
	\vertex [dot] (4) at(2.5,0){} [];
	\vertex (5) at(3.5,+1);
	\vertex (6) at(3.5,-1);
	\vertex (7) at(4.25,0){\(+\)};
	\vertex (8) at(7,-1);
	\vertex (9) at(5,-1);
	\vertex [dot] (10) at(6,0){} [];
	\vertex [dot] (11) at(7.5,0){} [];
	\vertex (12) at(6.5,1);
	\vertex (13) at(8.5,+1);

	\diagram {
		(2) -- [dashed,blue] (3),
		(1) --[fermion] (3) --[fermion](4) --[fermion](5),
		(4) -- [dashed,blue] (6),
	};
	
	\diagram {
		(8) -- [dashed,blue] (10),
		(9) --[fermion] (10) --[fermion](11) --[fermion](13),
		(11) -- [dashed,blue] (12),
	};
	\end{feynman}
	\end{tikzpicture}
	\caption{the tree diagram of interaction $\Psi+\Phi_m\rightarrow\Psi+\Phi_m$}
	\label{fig}
\end{figure}
$$\begin{array}{l}
\displaystyle {\cal S}^{(2)}=
\dfrac{i{\cal G}^2}{\hslash^2}\int\int d\mu(x_1)d\mu(x_2)
\left(
\gamma^4\cancel{B}^\top H\cancel{x}_1\;\boldsymbol{S}(x_1,x_2) \gamma^4\;H\cancel{x}_2\cancel{B}^\top
\right)_{ir}
\\
\\
\displaystyle
\qquad\qquad\qquad \times\left[
\bar{\Psi}^{\;-}_i(x_1)\Phi_m^-(x_1)\Phi_m^+(x_2) \Psi^+_r(x_2)+\bar{\Psi}^{\;-}_i(x_1)\Phi_m^-(x_2)\Phi_m^+(x_1) \Psi^+_r(x_2)
\right].
\end{array}
$$
By using \eqref{psi expansion} and \eqref{mmc sc-fi operator} and by notice that the incoming state ($\left|i\right>$) and outgoing state ($\left|f\right>$) each include a quanta of spinor field and a quanta of \textit{mmc} scalar field, one can obtain ${\cal S}_{fi}^{(2)}=\left<f\right|{\cal S}^{(2)}\left|i\right>$ as:
\b\label{s2}
\begin{array}{l}
	\displaystyle {\cal S}^{(2)}_{fi}=i\dfrac{{\cal G}^2}{\hslash^2}\;   \boldsymbol{N}_\Psi\boldsymbol{N}'_\Psi
	\boldsymbol{N}_{\Phi_m}\boldsymbol{N}'_{\Phi_m}
	\sum_{l,l'=1}^{5}\;\sum_{\sigma,\sigma'=-\frac{1}{2}\frac{1}{2}}\int\int\ud\mu(x_1)\ud\mu(x_2)	
	\\
	\\
	\displaystyle \times
	\bar{\cal{U}} ({\xi}'_p,\sigma')\gamma^4\;\cancel{B}^\top \;H\cancel{x}_1\;\boldsymbol{S}(x_1,x_2)\gamma^4\;H\cancel{x}_2\;\cancel{B}^\top\;{\cal{U}}({\xi}_p,\sigma) (Hx_1.{\xi}_p)^{-2+i\nu}
	(Hx_2.\xi'_p)^{-2-i\nu} \;
	\\
	\\
	\displaystyle \times\left\{
	\left[
	\left(HA^{(l)}\cdot \partial_1^\top+2H^3A^{(l)}\cdot x_1\right)(Hx_1\cdot\xi_k)^{-1}
	\right]
	\left[
	\left(HA^{(l')}\cdot \partial_2^\top+2H^3A^{(l')}\cdot x_2\right)(Hx_2\cdot\xi_k')^{-2}
	\right]
	\right.
	\\
	\\
	\displaystyle \left.
	+\left[
	\left(HA^{(l')}\cdot \partial_1^\top+2H^3A^{(l')}\cdot x_1\right)(Hx_1\cdot\xi_k')^{-2}
	\right]
	\left[
	\left(HA^{(l)}\cdot \partial_2^\top+2H^3A^{(l)}\cdot x_2\right)(Hx_2\cdot\xi_k)^{-1}
	\right]
	\right\}
	\\
	\\
	=\displaystyle i\dfrac{{\cal G}^2}{\hslash^2}\;   \boldsymbol{N}_\Psi\boldsymbol{N}'_\Psi
	\boldsymbol{N}_{\Phi_m}\boldsymbol{N}'_{\Phi_m}
	\sum_{l,l'=1}^{5}\;\sum_{\sigma,\sigma'=-\frac{1}{2}\frac{1}{2}}\int\int\ud\mu(x_1)\ud\mu(x_2)	
	\\
	\\
	\displaystyle \times
	\bar{\cal{U}} ({\xi}'_p,\sigma')\gamma^4\;\cancel{B}^\top \;H\cancel{x}_1\;\boldsymbol{S}(x_1,x_2)\gamma^4\;H\cancel{x}_2\;\cancel{B}^\top\;{\cal{U}}({\xi}_p,\sigma) (Hx_1.{\xi}_p)^{-2+i\nu}
	(Hx_2.\xi'_p)^{-2-i\nu} \;
	\\
	\\
	\displaystyle \times\left\{
	\left[
	H^2 (A^{(l)}\cdot x_1)(Hx_1\cdot\xi_k)^{-1}
	-
	H^2(A^{(l)}\cdot\xi_k)(Hx_1\cdot\xi_k)^{-2}\right]
	\left[
	-2H^2(A^{(l')}\cdot\xi'_k)(Hx_2\cdot\xi'_k)^{-3}
	\right]
	\right.
	\\
	\\
	\displaystyle \left.
	+\left[
	H^2(A^{(l)}\cdot x_2)(Hx_2\cdot\xi_k)^{-1}
	-
	H^2(A^{(l)}\cdot\xi_k)(Hx_2\cdot\xi_k)^{-2}
	\right]
	\left[
	-2H^2(A^{(l')}\cdot\xi'_k)(Hx_1\cdot\xi'_k)^{-3}
	\right]
	\right\}.
\end{array}
\e
This integrals can be calculated directly by numerical methods but here the behavior of ${\cal S}$-matrix in null curvature limit is of interest for us.

\section{the null curvature limit}\label{flat}
By being enlarged dS hyperboloid radius $R=H^{-1}$, the curvature of space time decrease and space-time become almost flat. In this null curvature limit ($R\rightarrow\infty$ or equivalently $H\rightarrow0$), the {\it mcc} scalar field and spinor field become their counterpart in the Minkowski space \cite{bagamota,brgamo,brmo}:
\b\label{psi flat limit}
\lim_{H\rightarrow 0}\Psi(x)=\psi(X)
\;,\;\;\;
\lim_{H\rightarrow 0}\bar{\Psi}(x)\gamma^4=\bar{\psi}(X), \;\;\; \lim_{H\rightarrow 0} \Phi_c(x)=\phi(X).
\e
The ambient space coordinates can be related with intrinsic coordinates in way:
 \b \label{flat cordinates}
x^\alpha= \left(
H^{-1}\sinh(HX^0),\;
H^{-1}\dfrac{\overrightarrow{X}}{\lVert\overrightarrow{X}\lVert}\cosh(HX^0)\sin(H\lVert\overrightarrow{X}\lVert),\;
H^{-1}\cosh(HX^0)\cos(H\lVert\overrightarrow{X}\lVert)\right),\e
then it is easy to show that:
$$
\lim_{H\rightarrow 0}H\barra{x}=-\gamma^4.\
$$
As it is discussed, the $A_\alpha$ and $B_\alpha$ are arbitrary constants then one can suppose the $A^4$ and $B^4$ are from order $H$. So the null curvature limit of $\barra{B}^\top$ is \cite{ahjata2019}
\b \label{B flat limit}
\lim_{H\rightarrow 0}\barra B^{\top} =-B_{\mu}\gamma^{'\mu}\gamma^4=-\barra{B}\gamma^4\;;\;\;\;\;\;\mu =0,1,2,3.\e
The null curvature limit of spinor two-point function is \cite{photon decaying}
\b\label{sp-fi 2point function flat limit}
\lim_{H \rightarrow 0}{\bf S}(x_1,x_2)\gamma^4=S(X_1,X_2)=\dfrac{1}{(2\pi)^4}\int\ud^4 q e^{iq\cdot(X_1-X_2)}\dfrac{\barra{q}+m}{q^2-m^2+i\epsilon}.
\e
Thus the null curvature of \eqref{s2} is:
$$
\begin{array}{l}
\displaystyle \lim_{H\rightarrow 0}{\cal S}^{(2)}_{fi}=-iH^2\dfrac{{\cal G}^2}{\hslash^2(2\pi)^4}\;   \boldsymbol{N}_\Psi\boldsymbol{N}'_\Psi
\boldsymbol{N}_{\Phi_m}\boldsymbol{N}'_{\Phi_m}
\sum_{l,l'=1}^{5}\;\sum_{\sigma,\sigma'=-\frac{1}{2}\frac{1}{2}}\int\int\ud^4X_1\ud^4X_2	
\\
\\
\displaystyle \times
\bar{\boldsymbol{u}} (p',\sigma')\;\cancel{B} \;
\left(\int\ud^4qe^{iq\cdot(X_1-X_2)}\dfrac{\barra{q}+m}{q^2-m^2+i\epsilon}\right)
\;\cancel{B}\;{\boldsymbol{u}}(p,\sigma) e^{ip\cdot X_1}e^{-ip'\cdot X_2}
\\
\\
\displaystyle \times\left\{
\left(A^{(l)}\cdot \partial\;e^{-ik\cdot X_1}\right)
\left(A^{(l')}\cdot \partial\;e^{ik'\cdot X_2}\right)
+
\left(A^{(l')}\cdot \partial\;e^{-ik'\cdot X_1}\right)
\left(A^{(l)}\cdot \partial\;e^{ik\cdot X_2}\right)
\right\}.
\end{array}
$$
By using definition of delta function in flat space the $\displaystyle\lim_{H\rightarrow 0}{\cal S}^{(2)}_{fi}$ is:
$$
\begin{array}{l}
\displaystyle\lim_{H\rightarrow 0}{\cal S}^{(2)}_{fi}=-i(2\pi)^4H^2\dfrac{{\cal G}^2}{\hslash^2}\;
\boldsymbol{N}_\Psi\boldsymbol{N}'_\Psi
\boldsymbol{N}_{\Phi_m}\boldsymbol{N}'_{\Phi_m}
\displaystyle\sum_{l,l'=1}^{5}\;\sum_{\sigma,\sigma'=-\frac{1}{2}\frac{1}{2}}
\left(A^{(l)}\cdot k\right)
\left(A^{(l')}\cdot k'\right)
\int\ud^4q
\\
\\
\displaystyle
\bar{\boldsymbol{u}} (p',\sigma')\;\cancel{B}\;
\dfrac{\barra{q}+m}{q^2-m^2+i\tau} \;\cancel{B}\;{\boldsymbol{u}}(p,\sigma)\;
\left\{
\delta^4(p+k-q)\delta^4(-p'-k'+q)
+
\delta^4(-p'+k+q)\delta^4(p-k'-q)
\right\}.
\end{array}
$$
The $A_\alpha$ satisfies this condition $\displaystyle\sum_{l,l'=1}^5 A_\alpha^{(l)}A_\beta^{(l')}=\eta_{\alpha\beta}$ \cite{77}. Given that this property of $A_\alpha$ and the delta function properties one can obtained the $\displaystyle\lim_{H\rightarrow 0}{\cal S}^{(2)}_{fi}$ as:
$$
\begin{array}{l}
\displaystyle\lim_{H\rightarrow 0}{\cal S}^{(2)}_{fi}=-i(2\pi)^4H^2\dfrac{{\cal G}^2}{\hslash^2}\;\left(k\cdot k'\right)\;
\boldsymbol{N}_\Psi\boldsymbol{N}'_\Psi
\boldsymbol{N}_{\Phi_m}\boldsymbol{N}'_{\Phi_m}
\delta^4(p+k-p'-k')
\\
\\
\times\displaystyle\sum_{\sigma,\sigma'=-\frac{1}{2}\frac{1}{2}}\bar{\boldsymbol{u}} (p',\sigma') \;\cancel{B}
\left[
\dfrac{\barra{p}+\barra{k}+m}{(p+k)^2-m^2+i\tau}
+
\dfrac{\barra{p}-\barra{k}'+m}{(p-k')^2-m^2+i\tau}
\right]
\;\cancel{B}\;\boldsymbol{u}(p,\sigma),
\end{array}
$$
it is clear that because of factor $H^2$ this term in null curvature limit vanished. Of course, this is expected result, because the null curvature limit of \textit{mmc} scalar field and thus the interaction Lagrangian \eqref{higgs spinor lagrangian} is zero. It can be showed by notice to the \eqref{magic}.
In other hand, the dS-spinor local transformation \eqref{transformation} in null curvature limit become constant phase transformation, which does not lead to any interaction.

\section{conclusion}\label{Conclusion}
The spinor-{\it mmc} scalar interaction $(\Psi+\Phi_m\rightarrow\Psi+\Phi_m)$ is investigated in ambient space formalism of dS universe. Investigating the interaction between the spinor field with \textit{mmc} scalar field shows that in null curvature limit, the local phase transformation of the dS-spinor field become a constant phase transformation so the interaction disappeared and then it is seen the interaction Lagrangian and two-point function vanished in this limit. In fact, in dS ambient space, the incoming spinor interacts with incoming \textit{mmc} scalar field and leads to outgoing spinor and \textit{mmc} scalar fields. It is clear that the incoming and outgoing spinor fields are different. But in null curvature limit the \textit{mmc} scalar field disappeared and the incoming and outgoing spinor are the same and no interaction occurred. In other word, vanishing of the curvature leads to disappearing of \textit{mmc} scalar field and its interaction with spinor field. This leads us to the conclusion that the \textit{mmc} scalar field can be due to the space-time curvature and thus it is expected that this field may be a part of a gravitational field.
\\
\\
{\bf{Acknowledgments}}:  The author wishes to express his particular thanks to  Prof. M. V. Takook for useful guidances and consultations and also to M. Dehghani,  M. Rastiveis, R. Raziani and S. Tehrani-Nasab for useful discussions.

\end{document}